\documentclass[prl,floats,aps,showpacs,twocolumn]{revtex4}
\usepackage{graphicx}
\usepackage{natbib}
\usepackage[usenames]{color}

\newcommand{\beq}{\begin{equation}} 
\newcommand{\eeq}{\end{equation}} 
\newcommand{\beqa}{\begin{eqnarray}} 
\newcommand{\eeqa}{\end{eqnarray}}

\def\mo{M$_\odot$}

\def\Dwa{$\,$\uppercase\expandafter{\romannumeral5}$\,$}

\def\sless{\lower2pt\hbox{$\buildrel {\scriptstyle <}
   \over {\scriptstyle\sim}$}}

\def\sgreat{\lower2pt\hbox{$\buildrel {\scriptstyle >}
   \over {\scriptstyle\sim}$}}
\def\sharpnull#1{}
\def\aa{Astron. Astrophys.}
\def\aap{Astron. Astrophys.}
\def\apj{Astrophys. J.}
\def\apjl{Astrophys. J. Lett.}
\def\apjs{Astrophys. J. Suppl. Ser. }
\def\mnras{Mon. Not. R. Astron. Soc.}

\begin{document}

\title{A New Mechanism for Gravitational Wave Emission in Core-Collapse Supernovae}

\author{Christian D. Ott$^{(1)}$, Adam Burrows$^{(2)}$,
Luc Dessart$^{(2)}$, Eli Livne$^{(3)}$}

\affiliation{$^{(1)}$Max-Planck-Institut f\"{u}r Gravitationsphysik,
Albert-Einstein-Institut, Potsdam, Germany; cott@aei.mpg.de}
\affiliation{$^{(2)}$Department of Astronomy and Steward Observatory, 
The University of Arizona, Tucson, AZ 85721;
burrows@zenith.as.arizona.edu,luc@as.arizona.edu}
\affiliation{$^{(3)}$Racah Institute of Physics, The Hebrew University,
Jerusalem, Israel; eli@frodo.fiz.huji.ac.il}

\begin{abstract}
We present a new theory for the gravitational wave signatures 
of core-collapse supernovae. Previous studies identified 
axisymmetric rotating core collapse, core bounce, postbounce convection, 
and anisotropic neutrino emission as the primary processes 
and phases for the radiation of gravitational waves.
Our results, which are based on axisymmetric, Newtonian
radiation-hydrodynamics supernova simulations (Burrows et al.\ 2006), 
indicate that the dominant emission process of gravitational
waves in core-collapse supernovae may be the oscillations of the
protoneutron star core. The oscillations are predominantly of
g-mode character, are excited hundreds of milliseconds after bounce, and
typically last for several hundred milliseconds. Our results
suggest that even nonrotating core-collapse supernovae should be visible
to current LIGO-class detectors throughout the Galaxy, and depending on progenitor
structure, possibly out to Megaparsec distances.
\end{abstract}


\maketitle

Ever since physicists began to think about gravitational
wave detection, core-collapse supernovae have been regarded
as prime sources. Traditionally, model calculations
estimating the supernova wave signature have focussed on the
gravitational-wave emission from rotating iron core collapse 
and core bounce (see, e.g., \cite{ott:04}). 
Recent results from stellar evolutionary
calculations \cite{heger:05,hirschi:04} and neutron star birth spin
estimates \cite{ott:06} indicate that presupernova
stellar iron cores may rotate much more slowly 
than previously assumed and that the asphericity during collapse
and bounce due to rotation is not generally great enough to produce a sizable
time-varying, wave-emitting mass-quadrupole moment. In fact, gravitational
radiation from large-scale postbounce convection and anisotropic neutrino 
emission are likely to exceed the bounce signal of such slowly rotating
supernova cores \cite{jankamueller:96,bh:96,ott:04,mueller:04}.

The first-generation LIGO-class detectors are now operating at design 
sensitivity and an international network of observatories, including 
LIGO, GEO600, VIRGO and TAMA, is on-line. 
Gravitational waves detected from a supernova can provide 
us with ``live'' dynamical 
information from the supernova core, complementing the supernova neutrino pulse
as the only other immediate carriers of information from deep inside the star. 
Using signal-processing techniques operating on a large
set of theoretical templates, it will be possible to extract 
supernova physics from a sufficiently strong signal 
\cite{summerscales:06}.

In recent simulations, Burrows et al.~\cite{burrows:06}
have observed that protoneutron star (PNS) core 
g-modes are excited by turbulence and by accretion downstreams 
through the unstable and highly-deformed stalled supernova shock 
(undergoing the Standing-Accretion-Shock Instability [SASI]
\cite{foglizzo:01,blondin:03,burrows:06,buras:06b})
at postbounce times of many hundreds of milliseconds.
The oscillations damp by the emission of strong sound waves
and do not ebb until accretion subsides. In this way the core 
g-modes act as transducers for the conversion of accretion 
gravitational energy into acoustic power that is deposited
in the supernova mantle and, as proposed by 
Burrows et al.~\cite{burrows:06}, may be sufficient to
drive an explosion.  Most easily excited is the 
fundamental $\ell$=1 core g-mode, but higher-order eigenmodes 
and, through nonlinear effects, harmonics of eigenmodes with 
complicated spatial structures, emerge at later times.

In this letter, we consider the intruiging possibility of
the emission of strong gravitational waves from the quadrupole
spatial components of the PNS core oscillations. We obtain 
new estimates for the gravitational-wave signature of core-collapse 
supernovae. With three different presupernova stellar models, 
we have performed the longest to date 2D Newtonian 
radiation-hydrodynamics supernova simulations.
We find that the gravitational waves from the quadrupole components 
of the core oscillations dominate the total wave signature 
in duration, maximum strain, and total energy emission by one to 
several orders of magnitude. 
We have also discovered 
an approximate progenitor dependence: more massive iron 
cores may experience higher frequency, higher 
amplitude oscillations, and, hence, more energetic 
gravitational-wave emission.

\emph{Method and Initial Models.}
We carry out our axisymmetric calculations 
with the VULCAN/2D code in the multi-group, flux-limited diffusion 
approximation \cite{livne:04,walder:05,burrows:06,dessart:06}. 
The computational grid is comprised of 120 angular
zones on a full hemisphere and 160 logarithmically-spaced, 
centrally-refined radial zones from 30~km out to 5000~km. The
innermost 30~km is covered by a Cartesian region that is deformed 
to smoothly match the polar grid at the transition radius \cite{ott:04}.
A resolution test with 50\% more angular and radial zones did not reveal
major differences in global integral observables such as the gravitational-wave
strain. We employ the Shen equation of state \cite{shen:98}.

We explore three models in this study. Model s11WW is the 11-\mo\
(Zero-Age Main Sequence [ZAMS]) presupernova model of 
Woosley~\&~Weaver \cite{ww:95} without rotation. 
Model s25WW is nonrotating as well, but is the  
25-\mo\ progenitor from the same study. Model m15b6 corresponds to the
15-\mo\ progenitor model of Heger et al. \cite{heger:05} which was evolved
with a 1D prescription for rotation and magnetic-field-driven
angular momentum redistribution. We map this model onto
our 2D grid under the assumption of constant rotation on
cylinders. It has a precollapse ratio of rotational kinetic energy to
gravitational potential energy, $\beta$~=~T/$|$W$|$, of 
$\sim$1$\times$10$^{-3}$\%. This value is one to two orders of magnitude smaller
than in previous models~ (e.g., \cite{harry:02b,ott:04,mueller:04}), but yields
a PNS consistent with neutron star birth spin estimates~\cite{ott:06}.

We extract gravitational waves from the mass motions via 
the quadrupole formula as described in \cite{finnevans:90,jm:97,ott:04}. 
In addition, we estimate the gravitational-wave emission by anisotropic neutrino
radiation with the formalism introduced by \cite{epstein:78} and 
concretized in \cite{bh:96,jm:97}.

\begin{figure}
\includegraphics[width=7cm]{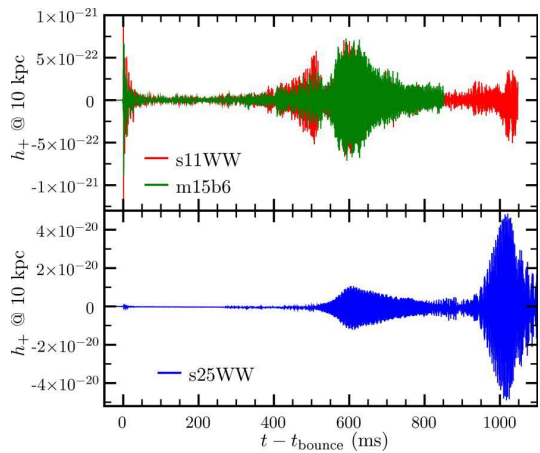}
\caption{Dimensionless gravitational-wave strain h$_+$ along the equator at a
distance of 10 kpc. Note that the range of h$_+$ in the lower panel is almost 50 
times wider than that of the top panel.
\label{fig:waveform}}
\end{figure}

\emph{Results}. Figure~\ref{fig:waveform} depicts the quadrupole gravitational
wave strain h$_+$ as emitted by mass motions scaled to a source distance of 
10~kiloparsecs (kpc). In the top panel, we superpose the waveforms of 
models s11WW and m15b6. Despite the presence of some rotation in the latter 
and its greater ZAMS mass, the two models have very similar precollapse stellar 
structures~\cite{ww:95,heger:05,ott:06}. This is reflected in the very similar 
shapes of their waveforms. Even though s11WW is not rotating, a bounce burst 
strain of $\sim$1.3$\times$10$^{-21}$ (@ 10 kpc) is 
present in our numerical model. The first one to two milliseconds of this burst
are the imprint of the transition in grid geometry from the outer polar to the inner
Cartesian grids which generates a time-varying quadrupole moment at core bounce. 
It also induces initial perturbations for 
vortical motion in the Ledoux-unstable regions behind the expanding shock that 
sets in almost immediately and with a perhaps too fast initial growth rate after core 
bounce. The amount of rotational energy in m15b6's core (at bounce, $\beta$\sless\ 0.02\% 
and at the end, $\beta$$\sim$0.08\%) is too small to have a large  
influence on the core dynamics and, thus, on the waveform, except to slightly 
stabilize the aspherical fluid motion at and shortly after bounce. 
In both models, until about $\sim$250~ms after bounce the 
physical waveform is dominated by convective motions in the  
PNS and in the post-shock region. As the 
SASI \cite{foglizzo:01,blondin:03,burrows:06,buras:06b}) becomes vigorous 
and leads to global deformation of the standing shock (see, e.g., \cite{burrows:06}) 
the wave emission from the post-shock flow increases.

\begin{figure}
\includegraphics[width=7cm]{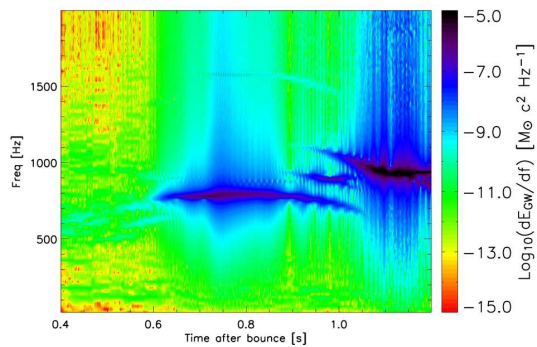}
\caption{
Frequency-time evolution of model s25WW's gravitational-wave energy spectrum
($dE_{\mathrm{GW}}$/$df$, \cite{jm:97}) computed with a 50 ms sampling interval.
\label{fig:timefreq}}
\end{figure}

As described in Burrows et al.~\cite{burrows:06} the fundamental core g-mode 
($\ell$=1,$f$$\sim$330 Hz in s11WW and m15b6) is excited by turbulence and accretion. 
It grows strong around $\sim$400~ms after bounce and starts 
transferring energy to the harmonic at 2$f$ through nonlinear effects. This is 
reflected in the rise of s11WW's gravitational-wave strain around that time.
h$_+$ reaches a local maximum, then quickly decays to about one-third that amplitude, 
only to pick up again after some tens of milliseconds, rising to even higher 
amplitudes (a maximum of
$\sim$7$\times$10$^{-22}$ [@ 10 kpc]), followed by a quasi-exponential 
decay with a $\sim$100~ms $e$-folding time. We attribute the 
gravitational-wave emission in these two `humps' to the quadrupole 
spatial component of the 2$f$ harmonic of the $\ell$=1 core g-mode. 
s11WW's gravitational-wave energy spectrum exhibits prominent emission 
in a band around $\sim$650~Hz. 
A frequency analysis shows that the harmonic appears first at a frequency of $\sim$590~Hz,
which increases over 200~ms to a maximum of about 680~Hz, and then continuously
decreases to $\sim$500~Hz at the end of the simulation. In this way, the gravitational-wave
emitting component exactly mirrors the behavior of the $\ell$=1 g-mode which goes through
the same phases \cite{burrows:06,burrows:06b}. This behavior is qualitatively 
consistent with the g-mode frequency evolution in PNSs presented 
in \cite{ferrari:03}, who used general relativity and obtained more compact 
PNSs with higher-frequency modes. Interestingly, we find that the time of the 
first lull in the wave emission ($\sim$500 ms) coincides with the point in time
at which the core-oscillation mode reaches its maximum frequency and its 
frequency derivative is zero~\cite{burrows:06b}. Even though model s11WW 
begins to explode around 550~ms, its core oscillation does not subside 
immediately \cite{burrows:06}. 

\begin{footnotesize}
\begin{center}
{\sc Table 1: Model Summary} \\
\vspace{0.1cm}
\begin{tabular}{lccccc}
\hline
\hline
\vspace{0.1cm}

Model & $\Delta$t$^{a}$ &$|$h$_{+,\mathrm{max}}|^b$ 
                    & h$_{\mathrm{char,max}}^{b,c}$ 
                    & f(h$_{\mathrm{char,max}}$) 
                    & E$_\mathrm{GW}^{d}$\\
&(ms)&(10$^{-21}$)&(10$^{-21}$)&(Hz)&(10$^{-7}$ \mo c$^2$)\\
\hline
s11WW&1045&1.3&22.8&654&0.16\\
s25WW&1110&50.0&2514.3&937&824.28\\
m15b6&927.2&1.2&19.3&660&0.14\\
\hline
\end{tabular}
\end{center}
$^a$\, time between bounce and the end of the simulation\\
$^b$\, at 10 kpc \hspace{.2cm} $^c$\, see Eq. (5.3) of \cite{flanhughes:98}.
$^d$\, see Eq. (12) of \cite{jm:97}
\end{footnotesize}
\vskip.1cm

\begin{figure}
\includegraphics[width=6.5cm]{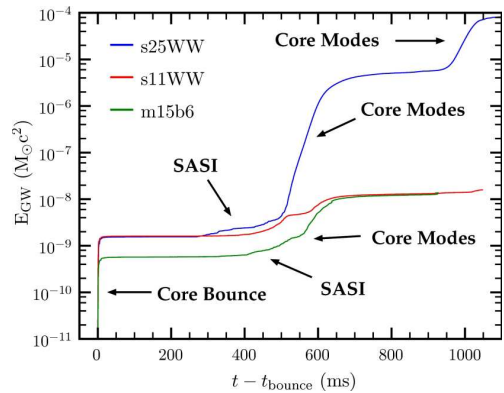}
\caption{
Integrated energy emitted in gravitational waves.
\label{fig:egw}}
\vskip-.4cm
\end{figure}

The lower panel of Fig.~\ref{fig:waveform} displays model s25WW's waveform. 
s25WW's precollapse stellar structure is significantly different from 
those of s11WW and m15b6. Most importantly, its iron core is more massive (1.92\mo\ vs.
1.37\mo\ and 1.47\mo, respectively) and more extended. These differences in progenitor
structure lead to different postbounce evolution for model s25WW. Its initial shock
radius is significantly smaller and the SASI becomes vigorous some 100~ms later
than in the two other models \cite{burrows:06c}. Figure~\ref{fig:timefreq} shows a
frequency-time plot of s25WW's gravitational-wave energy spectrum starting at
400~ms after bounce. The first burst of gravitational waves, starting at about
500~ms and slowly fading afterwards (Fig.~\ref{fig:waveform}), is centered about 
$\sim$800~Hz. In this model, the $\ell$=1 fluid mode, which dominates the dynamics 
at that time, is centered at $\sim$400~Hz and as for models s11WW and m15b6, 
we identify the wave-emitting component as part of the harmonic at 2$f$ of the former. 
At $\sim$900~ms after bounce, much stronger waves begin
to be emitted through the excitation of an $\ell$=2 core 
eigenmode (Fig.~\ref{fig:timefreq}; at 
that time $f\sim$950~Hz). It reaches a maximum 
strain of $\sim$5$\times$10$^{-20}$ (@ 10 kpc) and lasts for at least 200~ms, 
emitting a total of close to $10^{-4}$ \mo\ c$^2$  ($\simeq$1.8$\times$10$^{50}$ erg!) 
in gravitational waves (Fig.~\ref{fig:egw}). 

For the three models considered here we find the contributions of anisotropic
neutrino emission to be completely negligible compared with those of the core oscillations.
Extracting the neutrino-flux anisotropies at an observer radius of 200~km, 
well outside the neutrinospheres, and extrapolating to 10~kpc distance we find
maximum strains of 5.5$\times$10$^{-23}$, 2.6$\times$10$^{-23}$, and 
1.3$\times$10$^{-23}$ for s25WW, s11WW, and m15b6, respectively. These numbers
are also up to five times smaller than previous estimates obtained with codes that
perform neutrino transport along rays, which tends to overemphasize anisotropies 
\cite{bh:96,jm:97,mueller:04}, while our flux-limited diffusion approach tends to
smooth them out slightly. Due to their low-frequency character, gravitational
waves from neutrinos may not be observable by 
Earth-based detectors

\emph{Summary}. 
We have derived and analyzed the gravitational-wave emissions in
long-term Newtonian 2D radiation-hydrodynamics 
supernova calculations for three different
progenitors, one of which includes angular momentum consistent with
the inferred rotation period of young pulsars. We find that the
gravitational-wave emissions of all models are dominated by the quadrupolar
components of the core g-mode oscillations first discovered by 
Burrows et al.\ \cite{burrows:06}. If the core oscillations 
seen in simulations do obtain, their gravitational-wave 
emission will exceed the emission from all previously 
considered emission processes, including rotational core bounce, 
convection, and anisotropic neutrino emission. Importantly, we point
out that the excitation of the core g-modes does not depend
on the particular explosion mechanism proposed by 
Burrows et al.\ \cite{burrows:06}, but is likely to be a generic
phenomenon induced by turbulence and accretion downstreams in
any suitably delayed explosion mechanism.

\begin{figure}
\includegraphics[width=6.8cm]{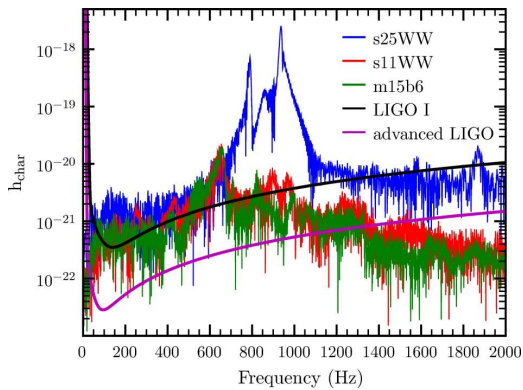}
\caption{
Characteristic strain spectra contrasted with initial and 
advanced LIGO (optimal) rms noise curves.\label{fig:ligospect}}
\end{figure}

The recent study of 
M\"uller et al.~\cite{mueller:04} suggests a total energy emission of 
3$\times$10$^{-9}$~\mo c$^2$ for a rotating model (initially about 3 times
as fast as our m15b6). For our nonrotating, low-mass model s11WW we find
about 5 times as much energy emission: 1.6$\times$10$^{-8}$~\mo c$^2$. 
For the slowly rotating, but otherwise very similar model m15b6, we calculate a 
total gravitational-wave energy of 1.4$\times$10$^{-8}$~\mo c$^2$
(Fig.~\ref{fig:egw} and Table 1). For the initial rotation rate and angular
momentum distribution present in m15b6, we do not find significant qualitative or
quantitative differences caused by such slow rotation. Model s25WW, due 
to its more massive iron core, higher postbounce accretion rates, and higher pulsation 
frequencies and amplitudes, emits an amazing 8.2$\times$10$^{-5}$~\mo c$^2$, 
almost one tenth of a typical supernova explosion energy. Clearly, radiation-reaction 
effects, not included in our present
work, might be significant for this model.
In Fig.~\ref{fig:ligospect}, we show the characteristic gravitational-wave
strain spectra of the three models, if located at 10~kpc.
$h_{\mathrm{char}}$ is defined by \cite{flanhughes:98} as 
$h_{\mathrm{char}}(f)$=$D^{-1}\sqrt{2 \pi^{-2} G c^{-3} dE_{\mathrm{GW}}/df}$, 
where $D$ is the distance to the source. This is a particularly useful measure, since it
incorporates the amount of energy radiated in a spectral interval $df$ around $f$.
In addition, we show the optimal rms noise strain ($h_{\mathrm{rms}}=\sqrt{f S(f)}$, 
$S(f)$ being the spectral strain sensitivity) of both LIGO~I 
and Advanced LIGO \cite{gust:99}. All h$_{\mathrm{char}}$ spectra peak 
strongly at the frequencies identified with the quadrupole
components of the core oscillations (Fig.~\ref{fig:timefreq}, between 600 and 1000 Hz, 
likely to be higher when general relativity is included), corroborating the narrow-band 
nature of the emission process. Given our results, we conclude that, if the core
oscillations observed in our simulations are generically excited in core-collapse
supernovae, even nonrotating supernovae of small to intermediate progenitor mass
should be observable by LIGO throughout the Milky Way and beyond. Massive progenitors
could be detectable out to $\sim$100 times greater distances and their prolonged and
extremely energetic core-oscillation wave signature might be the generic precursor of
stellar-mass black-hole formation.

We point out that the work presented here is based on 
simulations in 2D Newtonian gravity and only quadrupole 
wave emission has been considered.
General relativity is likely to increase the frequency of the PNS eigenmodes, 
but is unlikely to lead to qualitative differences. Fast rotation might 
lead to the partial stabilization of 
the post-shock convection, might affect the growth of core oscillations,  
and will likely lead to nonaxisymmetric rotational instabilities 
for $\beta$ \sgreat\ 8\% \cite{rotinst:05}.  In 3D, the temporal and spatial
mode and SASI structures may change.

\acknowledgments

We acknowledge help from and discussions with C. Meakin, J. Murphy,
H. Dimmelmeier, E. M\"uller, J. Pons, N. Stergioulas, L. Rezzolla, and E. Seidel. 
We acknowledge support from the Scientific Discovery 
through Advanced Computing (SciDAC) program of the DOE, grant number DE-FC02-01ER41184,
and from the NSF, grant number AST-0504947. 
E. L. acknowledges support from the Israel Science
Foundation under grant 805/04.
This research used resources of the National
Energy Research Scientific Computing Center, 
which is supported by the Office of Science of the U.S. Department of 
Energy under Contract No. DE-AC03-76SF00098.

\end{document}